\documentclass[graybox]{svmult}

\usepackage{mathptmx} 
\usepackage{helvet} 
\usepackage{courier} 
\usepackage{type1cm} 
\usepackage{makeidx} 
\usepackage{graphicx} 
\usepackage{multicol} 
\usepackage[bottom]{footmisc}

\usepackage{amsmath}
\usepackage{amsfonts}
\usepackage{graphicx} 

\makeindex 

\newcommand\coma{A}
\newcommand\aaa{V}
\newcommand\Hi{{\cal H}}
\newcommand\glf{\underline{\Sigma}}
\newcommand\pads{\mathcal{C}}

\newcommand\Sets{{\bf Sets}}
\newcommand\sss{\mathcal{S}}
\newcommand\stime{M}

\newcommand\bigaaa{\mathcal{B}(\Hi)}

\newcommand\be[1]{\begin{equation}\label{#1}}
\newcommand\ee{\end{equation}}

\def\mv{\mathcal{V}}
\def\mh{\mathcal{H}}

\begin{document}

\title*{Minimax context principle}
\author{Rom\`an Zapatrin\footnote{Dept. of Informatics, The State Russian museum, In\.zenernaya, 4, 191186, St.Petersburg, Russia, email: roman.zapatrin at gmail.com}}
\author{Rom\`an Zapatrin}
\institute{Rom\`an Zapatrin \at Dept. of Informatics, The State Russian museum, In\.zenernaya, 4, 191186, St.Petersburg, Russia, \email{roman.zapatrin@gmail.com}}
%
%
\maketitle

\abstract{I show how space-like structures emerge within the topos-based approach to quantum mechanics. With a physical system, or, more generally, with an operationalistic setup a {\emph context category} is associated being in fact an ordered collection of contexts. Each context, in turn, is associated with certain configuration space. The {\emph minimax context principle} is put forward. Its basic idea is that among various configuration spaces the `physical space' is the configuration space of a structureless point particle. In order to implement it, two order relations on contexts are introduced being analogs of inner and outer daseinisation of projectors. The proposed minimax context principle captures two characteristic features of physical space: maximal with respect to refining the accuracy, and minimal by getting rid of extra degrees of freedom.}

\section*{A Foreword}
\label{foreword}
The main goal of this paper is to explore new options, which are provided by topos approach to quantum mechanics. This approach was initially aimed to bring objectivity to quantum mechanics. Its further development made it more general, applicable to a broad class of operationalistic theories, and it was named ``Topos foundation for theories of physics'' \cite{id1,id2,id3}, denoted in the sequel by TFTP. In this essay I try to show that within TFTP we can describe how physical space is created as a result of measurements. Why `created' rather than `explored'? In brief, the reason is exactly the same as the reason why the value of the spin of a polarized particle emerges as a result of experiment; I illustrate it by a toy model of `topologimeter'. 

\medskip

Begin with a conventional paradigm: we are living in a physical spacetime $\stime$. The first step outwards is to state that the spacetime is something pre-existing and we are trying to measure it in whatever sense. `To measure' means to learn its structure. Moving towards operationalistic viewpoint, we adopt that our devices are not of absolute precision and, with respect to the idealistic spacetime manifold we are dealing with its partition -- some events become indistinguishable, this is called \emph{coarse-graining}. If we are given a manifold, all its coarse-grainings form a partially ordered set $\pads$ , and the initial spacetime $\stime$ is the maximal element of $\pads$, corresponding to ideal precision. 

On the other hand, we may explore many-particle systems, each such system has a configuration space which is, roughly speaking, like a Cartesian power of $\stime$. The configuration space (perhaps coarse-grained, TFTP is flexible enough) of $n$-particle system is embedded into that of $(n+1)$-particle system. We consider the set $\pads$ containing all available configuration spaces to those associated with multipartite systems and the resulting space now bears two partial orders: one associated with coarse-graining (called it precision order $\vdash$), and the other associated with ignoring extra degrees of freedom (called redundancy order $\rhd$). From this we observe how the initial spacetime is positioned among the available configuration spaces: it is maximal with respect to precision and minimal with respect to redundancy. This is how it looks in classical mechanics.

\medskip

However, in a more general operationalistic setting we do not consider the above mentioned configuration spaces as primary objects. One of the basic ingredients of TFTP is the notion of \emph{context category}. For a physical system, or, more general, for an operationalistic environment, the context category is a family of commutative subalgebras of observables treated as operationalistic `snapshots of reality'. Due to Ge'fand transform (which associates to a given commutative algebra $V$ a set $\glf(V)$ --- its spectrum), they are treated as factory of configuration spaces. In the classical case that we considered above, context category is the extended space $\pads$ with a natural ordering being simply the set inclusion. These order relations treated as arrows make $\pads$ category. However, there are two more partial orders on contexts, which are induced by daseinisation procedure, turning TFTP into a factory producing configuration spaces. Let us consider all this in more details restarting from the classical case. 

\section{Contexts and their supports in classical realm}

Suppose we are dealing with a classical physical system $\sss$, let $\stime$ be its configuration space. That means, each point of $\stime$ bears the information about the results of all queries addressed to $\sss$. Observables of $\sss$ are functions on $\stime$, denote the set of all observables by $\aaa$. The set $\aaa$ is a commutative algebra as its elements are functions, which are multiplied pointwise. We may, instead, consider $\aaa$ as a primary object, just a collection of elements, which can be added and multiplied by each other or by a number. The important result is that $\stime$ is recovered from $\aaa$, and this is the essence of Gel'fand transform: 
\be{glf}
\aaa \mapsto \glf(\aaa)
\ee
where $\glf(\aaa)$ is by definition the set of all multiplicative linear functionals on $\aaa$. So, given a commutative algebra, Gel'fand transform always return a set, which we interpret as a configuration space. In the language of TFTP approach, the algebra $\aaa$ is called \emph{context} and the resulting set $\glf(\aaa)$ is called the \emph{support} of the context $\aaa$. Configuration spaces are supports of contexts, that is why context category are `factories of configuration spaces'.

\section{Device resolution order} 

For various reasons we may consider different algebras $\aaa$ associated with the same system. First, we may set up certain threshold of accuracy, so that some measurements will be no longer available. That means, a smaller algebra $\aaa^\prime$ is considered being a subset of $\aaa$. Due to the duality, the associated configuration space $\glf(\aaa^\prime)$ is a quotient of $\glf(\aaa)$, it is called a \emph{coarse-graining} of $\glf(\aaa)$. If we consider a collection of subalgebras of $\aaa$, we have a partially ordered set with the greatest element $\aaa$, and dually, we have a family of coarse-grainings of $\glf(\aaa)$ ordered by projection, where $\glf(\aaa)$ itself is the greatest element. Let us call it \emph{resolution order}, it orders contexts by the resolution of available measuring devices, denote it 
\be{res}
\aaa^\prime \vdash \aaa \quad\Rightarrow\quad \glf(\aaa) \to \glf(\aaa^\prime)
\ee
The order $\vdash$ on context means that every statement (query) $Q$, which can be formulated in $\aaa^\prime$ can also be formulated in $\aaa$ and has the same truth value. 

\begin{figure}[h!]
\centering
\includegraphics[width=90mm]{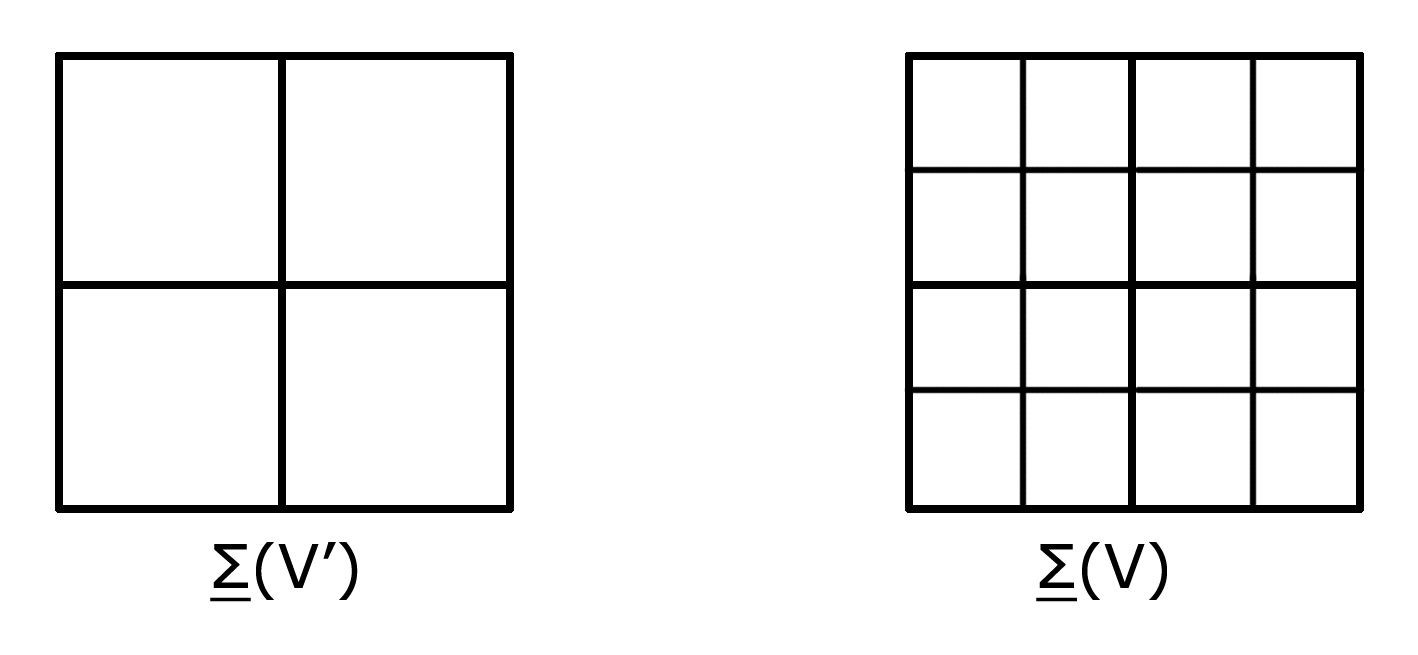}
\caption{Device resolution order (weaker $\vdash$ stronger)\label{DevResOrd}}
\end{figure}

\section{Redundancy order} 

On the other hand, a system may possess internal degrees of freedom, or may consist of several particles. In this case we may disregard some of the extra degrees of freedom within the initial algebra $\aaa$. The resulting algebra $\aaa'$ in this case is the quotient of the initial algebra $\aaa$. Dually, the appropriate configuration spaces are ordered by set inclusion
\be{red}
\aaa^\prime \rhd \aaa \quad\Rightarrow\quad \glf(\aaa) \subseteq \glf(\aaa^\prime)
\ee
So, the most simplified configuration space is associated with the minimal algebra with respect to passing to quotient. 

\begin{figure}[h!]
\centering
\includegraphics[width=90mm]{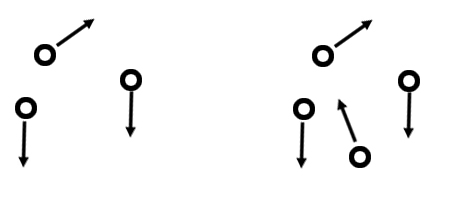}
\caption{Redundancy order (less redundant $\lhd$ more redundant)\label{DevResOrd}}
\end{figure}

Let us call the order \eqref{red} \emph{redundancy order}, it orders contexts by the possibility of getting rid of redundant degrees of freedom. In terms of queries, that means that each query formulated in $\aaa$ can be translated into a query in $\aaa^\prime$ by disregarding redundant data. 

\section{Minimax context principle for classical systems} 

Now let us figure out how the initial configuration space is positioned among all these spaces. It is the finest among coarse-grainings and in the same time it is the least informative. In terms of algebras of observables that means that the algebra related to what we could call `physical space' is maximal with respect to resolution order $\vdash$ \eqref{res} and minimal with respect to redundancy order $\rhd$ \eqref{red}. 

\medskip

There are two important observations. First, both orders are not a part of classical mechanics, they are imposed by the model. For instance, the redundancy order does not make difference between internal degrees of freedom of a single particle and multipartite system. Second, the resolution order is imposed by extra assumption about the accuracy of available devices, it directly reflects the operationalistic approach. As a consequence, even within the classical setting we have a variety of configuration spaces of the same system. 

At first sight, two orders are the same, both they are formulated as set inclusions. However, they are of different nature: the resolution order $\vdash$ is associated with the outer daseinisation, while the redundancy order is produced by the inner daseinisation, both are formulated in TFTP, let us dwell on it in a more detail. 

\section{TFTP and daseinisation} 

Nowadays topos approach to quantum mechanics is a well-developed paradigm. I will only outline its ingredients, which are relevant for this essay. For details the Reader is referred to a review \cite{floribrief} or lecture notes \cite{florifull}. 

One of the achievements of TFTP is merging the idea of realism with the mathematical machinery of quantum mechanics. The r\^ole of the set of states, or a generalized configuration space is played by a topos rather than by a set. I do not even want to provide the definition of topos here, only present it explicitly. This topos, the core ingredient of TFTP is formed as follows. Given a `big' in a sense quantum system with the state space $\mh$, the collection of all its \emph{abelian} subalgebras of the von Neumann algebra $\bigaaa$ of all bounded operators on $\Hi$ is called \emph{context category}. In fact, the category $\mv(\mh)$ is a partially ordered set ordered by set inclusion. This category is very important in the formulation of quantum theory in terms of topos theory. But effectively we need only the partial order on $\bigaaa$. Each context contains idempotent elements, they are referred to as \emph{queries}. 

The next important notion of TFTP is daseinisation. Given a query $P$ (a projection operator in quantum mechanics) and given a context $V$, the daseinisation aims to reconcile them. If $P$ belongs to the context, there is nothing to reconcile and the daseinisation procedure returns $P$ itself. If $P\not\in V$, then there are two kinds of daseinisations - outer and inner one.
\begin{itemize}
\item The outer daseinisation returns a minimal projector $\delta^o(\hat{P})_V \in V$, which contains $P$ -- in language terms, the most detailed query in $V$, which follows from $P$. So,
\[
\delta^o(\hat{P})_V = 
\bigvee \left\lbrace
Q\:\vert\: P\vdash Q
\right\rbrace
\]
\item The inner daseinisation, in contrast returns the maximal projector $\delta^i(\hat{P})_V \in V$, which is contained in $P$ -- in language terms, the least detailed query in $V$, from which $P$ follows. 
\[
\delta^i(\hat{P})_V = 
\bigwedge \left\lbrace
Q\:\vert\: Q\rhd P
\right\rbrace
\]
\end{itemize}

\section{Minimax principle in general TFTP operationalistic models}\label{minimaxgen}

The elements (objects, strictly speaking) of context category are commutative algebras. For them, the Gel'fand duality is considered. With each commutative subalgebra $\aaa$ of $\bigaaa$ its Gel'fand spectrum $\glf(\aaa)$ \eqref{glf} is associated. Then we proceed exactly in the same way as we did in the classical realm. But now, from the very beginning there is no `true' space underlying the whole scope of the observations. Therefore, each minimax context can be treated as a physical space: there is no indication within TFTP what is more and what is less `real', or physical. Being applied to quantum mechanical systems, the algebra $\bigaaa$ is the von Neumann algebra of bounded operators on the Hilbert space $\Hi$ associated with the system. The structure analogous to the state space of the overall system is the spectral presheaf $\mbox{\Sets}^{\mv(\mh)}$. 

The machinery itself, the basic ideas of TFTP are more general than just a reformulation of quantum mechanics. They may be applied to any operationalistic environment. In general, it looks like a dialog of an Observer with an External Environment -- whatever it be: a display, a control center or a storage of datasheets. Anyway, the methodology remains: the Observer inputs queries and then receives replies. After a series of queries an appropriate datasheet is formed. From it, the algebra of observables is inferred. The algebra of observables is in general a non-commutative algebra $\bigaaa$. To link it with spatial structures, we consider its commutative sub-algebras\footnote{A more general construction, employed in quantum gravity \cite{cqg}, exists for non-commutative algebras, where points are reconstructed as irreducible representations, which makes it possible to endow finite sets with non-trivial (that is, non-discrete as it always takes place for Gel'fand transform) topology. This is beyond consideration in this essay.}. I emphasize that it is not necessary to take all commutative sub-algebras into account. Only available ones, that is, generated by available observables subalgebras are considered. 

From the algebraic point of view, the central point of minimax principle is that two new partial orders are introduced on $\mv(\mh)$, each being weaker than the initial one. These are the resolution order \eqref{res} and the redundancy order \eqref{red}. 

\begin{figure}[h!]
\centering
\includegraphics[width=110mm]{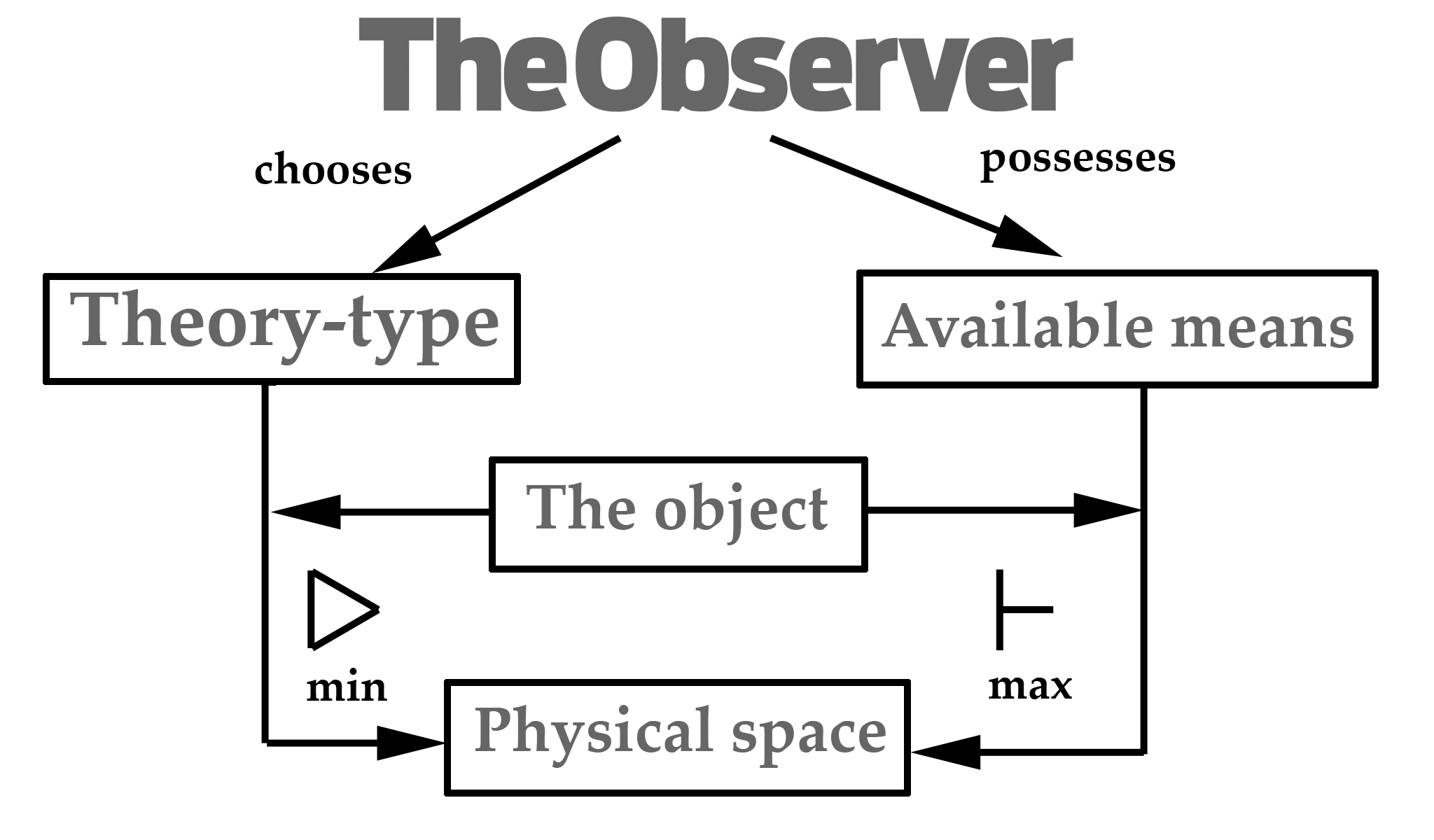}
\caption{The minimax principle \label{overflow}}
\end{figure}

\section{A toy model of `topologimeter'} 

Suppose we have an Observer, who is given a lot of data, yet unordered. Each query provides a datasheet of, say 10000 entries. The first step for the observer to somehow structure the data is to employ, say, factor analysis. Suppose it is done, and the result is that each datasheet can be represented as a $100\times 100$ table of numbers. There are very many such tables, and the observer, in order to simplify the model, approximates it by functions of two real variables, each datasheet is now treated as a function $f(x,y)$. 

From now on we are going to treat overall results as observations over particles on a configuration space. First of all, let us consider commutative algebra $\coma$ generated by the obtained functions $f(x,y)$. Then form the von Neumann algebra of matrices whose entries are the elements of $\coma$, supposed they are treated as available (as it was emphasized in section \ref{minimaxgen}.
\[
\aaa = 
\left\lbrace
\left(\begin{array}{cc}
a_{11}(x,y) & a_{12}(x,y)
\\
a_{21}(x,y) & a_{22}(x,y)
\end{array}\right)
\right\rbrace
\]
with $a_{21}=\overline{a_{12}}$. Define two its maximal commutative subalgebras. The first is 
\be{twolines}
\aaa_1 = 
\left\lbrace
\left(\begin{array}{cc}
f(y) &g(y)
\\
g(y) & f(y)
\end{array}\right)
\right\rbrace
\ee
The second is 
\be{linecirc}
\aaa_2 = 
\left\lbrace
\left.
\left(\begin{array}{cc}
p(x) & 0
\\
0 &q(x)
\end{array}\right)
\quad
\right\vert
\quad
p(x)=p(x+2\pi)
\right\rbrace
\ee
It is easy to check by direct calculation that both $\aaa_1,\aaa_2$ are commutative and maximal subalgebras of $\aaa$. Calculate the appropriate Gel'fand spaces for them. For $\aaa_1$ we have the disjoint sum of domains of the function $f$ and $g$, that is, two disjoint straight lines. For $\aaa_2$ we have the disjoint sum of a circle (because all functions $p(x)$ are periodical) and a line (since $q(x)$ has no restrictions). Speculating with this, we may state that the context $\aaa_1$ does not admit topology change, while $\aaa_2$ may be treated as `topologimeter'. 

\unitlength1mm
\thicklines
\[
\begin{array}{ccc}
\begin{picture}(40,40)
\put(1,1){\line(1,1){20}}
\put(1,14){\line(1,1){20}}
\end{picture}
& \qquad \qquad &
\begin{picture}(40,40)
\put(1,1){\line(1,1){20}}
\put(7,23){\circle{20}}
\end{picture}
\\
\glf(\aaa_1)
& &
\glf(\aaa_2)
\end{array}
\]

Note that the subalgebras \eqref{twolines},\eqref{linecirc} are maximal subalgebras of $\aaa$. However, $\aaa$ itself may be a subalgebra of a bigger algebra $\mathcal{A}$. Within this bigger algebra, $\aaa$ has the form 
\[
\aaa = 
\left\lbrace
\left(\begin{array}{cc}
a_{11}(x,y) & a_{12}(x,y)
\\
a_{21}(x,y) & a_{22}(x,y)
\end{array}\right)
\bigotimes
\mathbb{I}
\right\rbrace
\]
where $\mathbb{I}$ stands for the unit operator. Operationally that means that all extra degrees of freedom are swept away. 

\medskip

Why do I call it `topologimeter'? If we choose the context $\aaa_1$ and perform all measurements within it, we reconstruct the configuration space being a straight lie (more precisely, a disjoint sum of straight lines, but they have the same topology). So,  the result of a measurement within the context  $\aaa_1$ always produces a physical space with the topology of line. If, instead, we choose the context $\aaa_2$, then the situation changes. The resulting post-measurement state if associated either with the `top-left' subalgebra of periodic functions, or with a `bottom-right' subalgebra, whose Gel'fand space is a line. This means, that the result of a measurement within the context  $\aaa_2$ yields different physical spaces having either the topology of a line, or that of a circle. 

\section*{So what?} 

What I tried to do in this essay, is to present a framework based on the Topos foundation for theories of physics (TFTP) to treat physical space itself and its topology as observables, to demonstrate that, like the values of momentum or spin, they emerge in the act of measurement. For that, the formalism of TFTP was used. I introduce an additional \emph{minimax context principle}, which generalizes TFTP's daseinisation procedure from projectors to whole contexts. Loosely speaking, I describe a factory of configuration spaces and a procedure making happy those who wish to perceive the reality in terms of clocks and rulers. 

I consider these ideas vital, because, in the light of new technologies the very notion of experiment broadens, the bounds between real and virtual smear out and virtually emerging spaces are to greater and greater extent observed in experiment. This essay is a part of a general research program, inspired by the idea that the `real world' is gradually moving towards virtualization: a nowadays researcher is a miner of Big Data rather than a `locksmith'.

\end{document}